\documentclass[10pt,a4paper]{article}
\usepackage[utf8]{inputenc}
\usepackage[T1]{fontenc}

\usepackage{amsmath}
\usepackage{amsthm}

\usepackage{amssymb}
\usepackage{mathrsfs}
\usepackage{eucal}[mathscr]
\usepackage{mathtools}
\usepackage{float}
\usepackage[pdftex]{graphicx}

\usepackage{supertabular}
\usepackage{multirow}
\usepackage{cancel}
\PassOptionsToPackage{hyphens}{url}
\usepackage{hyperref}

\numberwithin{equation}{section}

\def\Real{\mathbb{R}}
\def\ZZ{\mathbb{Z}}
\def\RR{\textbf{\textit{R}}}
\def\Gg{\textbf{\textit{G}}}

\def\RRic{\mathbb{\mathcal{R}}_{ic}}
\def\Rsc{\mathsf{R}}

\def\gm{\mathbf{g}} 

\def\eps{\varepsilon}
\def\ffi{\varphi}

\def\engie{\textsf{e}}
\def\Lcal{\mathcal{L}}

\def\WRRic{{}^{\scriptscriptstyle{W}}\mkern-6mu \RRic}

\def\WRR{{}^{\scriptscriptstyle{W}}\mkern-6mu \RR}
\def\WRsc{{}^{\scriptscriptstyle{W}}\mkern-4mu \Rsc}

\def\WGG{{}^{\scriptscriptstyle{W}}\mkern-6mu \textbf{\textit{G}}}

\def\Wnabla{{}^{\scriptscriptstyle{W\!}}\nabla}

\def\divpsi{\textrm{div}\,\psi_*}

\newcommand{\dlam}[1]{\tfrac{d{#1}}{d\lambda}}
\newcommand{\ddlam}[1]{\tfrac{d^2{#1}}{d\lambda^2}}

\newcommand{\dmu}[1]{\tfrac{d{#1}}{d\mu}}

\DeclareMathOperator{\Tr}{Tr}

\def \drawfigtwo#1#2#3{\centerline{\includegraphics*[scale=#3,clip=false]{#1}%
\includegraphics*[scale=#3,clip=false]{#2}}}

\begin{document}

\author{Michel Duneau \\
{\normalsize  Centre de Physique Th\'eorique, CNRS, Ecole polytechnique,}\\
{\normalsize  Institut Polytechnique de Paris, Palaiseau, France}} 

\title{Weyl locally integrable conformal gravity, rotation curves and cosmic filaments} 
\date{June 7, 2022}
\date{\today}

\maketitle


\begin{abstract}
Weyl conformal theory of gravity is an extension of Einstein theory of general relativity, in which metrics $\gm$ are 
associated to 1-forms $\psi$. A Weyl structure is defined by an equivalence class of pairs $(\gm,\psi)$ where the 
metrics $\gm$ are conformally equivalent and the 1-forms $\psi$ differ by exact 1-forms. 
The simplest non-trivial case assumes that the 1-forms $\psi$ are closed but non-exact, implying that 
the spacetime manifolds are not simply connected. The Einstein tensors computed with the Weyl connections have null divergence and they \textit{implicate} a cosmological term replacing the constant $\Lambda$ by a 
function of spacetime, and a shear stress tensor. 

A toy model based on the Schwarzschild metric is presented. The associated 1-form is locally proportional to $d\ffi$ 
in Schwarzschild coordinates $(ct,r,\theta,\ffi)$, with an expected effect on galactic rotation curves. 
Depending on initial conditions planar geodesics show almost constant total velocities regardless of $r$.
This also implies a singularity on the whole $z$ axis: In the free case $r_S=0$, induced spin effect in its neighbourhood 
are comparable to recent observations concerning cosmic filaments. 
\end{abstract}

\section{Introduction}
The Weyl conformal geometry has received a growing interest in the past years 
(\cite{MAEDER2019_2,SCHOLZ3,SCHOLZ2}), 
especially on the subject of galactic rotation curves 
(\cite{DELIDUMAN,EDERY2,MANNHEIM_1995,MANNHEIM_2006,MANNHEIM_2012} and references cited therein). 
It is an extension of Riemannian geometry where the Levi-Civita connection $\nabla$ 
associated to a metric $\gm$ is specified by the invariance condition $\nabla\gm=0$. 
The Weyl connection $\Wnabla$ is the unique torsion free connection which satisfies the weaker condition 
$\Wnabla_\xi\gm = -2\psi(\xi)\gm$, where $\psi$ is a 1-form associated to $\gm$. This property extends 
at once to an equivalence class of pairs $(\gm,\psi)$ where the metrics $\gm$ are conformally equivalent and the 1-forms 
$\psi$ differ by exact 1-forms. The case where the 1-forms $\psi$ are exact has been extensively studied as there always exists 
a compatible metric associated to a null 1-form, so that the Weyl connection is identical to the Levi-Civita connection for that 
me\-tric. It follows that all derived objects such as Riemann curvature tensors, Ricci tensors, scalar curvatures and Einstein tensors are identical for both connections. 
The general case of not closed 1-forms is much more complicated as the Ricci tensor resulting from 
$\Wnabla$ is no more symmetric. 

Section 2 recalls basic features of Weyl conformal manifolds and presents some of their particular properties when the 
1-forms of the pairs $(\gm,\psi)$ are locally integrable (closed and non-exact). In these cases the Weyl-Riemann curvature tensor $\WRR$, computed with $\Wnabla$, satisfies the same five properties as the  Riemann curvature tensors. 
The  Weyl-Einstein tensor $\WGG$ has zero divergence and differs 
form the (pseudo-)Riemannian ones by 2 terms: a \textit{non-constant} scalar function depending on $\psi$, in place of the  
optional cosmological constant $\Lambda$, and a shear stress tensor. Existence of closed non-exact 1-forms on a 
manifold $M$ implies topological restrictions on $M$ like the fact that $M$ is not simply connected. 
The principal outcome is the Weyl-Einstein tensor and the comparison with the Einstein tensor for a given metric $\gm$. 

In section 3 we consider the geodesics of a Weyl connection and their preferred parametrization depending on 
$(\gm,\psi)$ in the Weyl structure. The geodesics equations are expressed with the Leci-Civita connection of $\gm$ and reveal 
acceleration terms involving the 1-form $\psi$. 
By Poincar\'e lemma these terms disappear for Levi-Civita connections of metrics $\gm'$ such that $\psi'$ locally vanishes. 

Section 4 presents a simple example of a locally integrable Weyl structure where $\gm$ is the Schwarzschild metric. The associated 1-form $\psi$ is locally equal to $\eps\,d\ffi$ in Schwarzschild coordinates $(ct, r, \theta,\ffi)$ and  $\eps$ is a 
dimensionless constant.  
The rotational force induced by the 1-form is invariant with respect to $z$ translations. The angular momenta of geodesics are 
affine functions of proper time and, according to initial conditions, total planar velocity curves turn out to be 
almost flat as functions of $r$. In the neighbourhood of the $z$-axis and in the free case $r_S=0$, geodesics are accelerated or 
slowed down according to which side of the $z$-axis they are going. 
 
\section{Locally integrable Weyl structures}
In his papers \cite{WEYL1,WEYL2} of 1918 Hermann Weyl introduced an extension of the Riemannian geometry by weakening 
the invariance condition $\nabla\gm=0$ which is known to define a unique torsion free Levi-Civita connection $\nabla$ 
for a given metric $\gm$. 
In a Weyl manifold $M$ the connection, noted $\Wnabla$ in the following, is subject to the weaker condition 
$\Wnabla_\xi\gm=-2\psi(\xi)\gm$, where $\psi$ is a 1-form on $M$ associated to $\gm$, for any tangent vector $\xi$. 
It follows that for any regular function $\Omega$ on $M$, the conformally equivalent metric $\gm'=e^{2\Omega}\gm$ 
satisfies the similar equation $\Wnabla_\xi \gm'=-2\psi'(\xi)\,\gm'$ with the associated 1-form $\psi'=\psi - d\Omega$. 
This gauge transformation leads to the equivalence relation \footnote{Different conventions are used in the literature. 
We adopt here that of E. Scholz \cite{SCHOLZ1} and others.}between pairs $(\gm,\psi)$ (see \cite{FOLLAND,SCHOLZ3,SCHOLZ1,WEYL2}):

\small
\begin{align}
\label{Weylequival}
(\gm, \psi) \equiv (\gm', \psi') \iff 
\left\{
\begin{array}{lr}
\gm' = e^{2\Omega} \gm, & \Omega \in C^\infty(M),\\
\psi'= \psi + \phi, & \phi = - d\Omega  . 
\end{array}
\right.
\end{align}
\normalsize
By definition a Weyl structure on a manifold $M$ is an equivalence class of pairs $(\gm,\psi)$. 
Then, a unique conformally invariant Weyl connection $\Wnabla$ is specified by the two conditions 
\small
\begin{align}
\label{Wnablag}
\begin{split}
&\Wnabla_X \gm = -2 \psi(X) \gm, \\
&\Wnabla_XY-\Wnabla_YX = [X,Y],
\end{split}
\end{align}
\normalsize
for any vector fields $X$ and $Y$. The first condition holds true for all pairs $(\gm,\psi)$ in the Weyl structure. 
The second one states that the connection is torsion free. Existence and unicity of $\Wnabla$ follow from an adaptation 
of the Koszul formula.\\

The Weyl structure is said to be closed if $d\psi=0$ and exact if $\psi=d\chi$ for some scalar function $\chi$. 
Each property is satisfied for all pairs $(\gm',\psi')$ in the equivalence class of the Weyl structure. 
If $\psi=d\chi$ and $\gm'=e^{2\chi} \gm$ we find that $(\gm',0)$ belongs to the Weyl structure. Caldebank et al. \cite{CALDERBANK} 
pointed out that the Poincar\'e lemma implies that a closed \emph{non-exact} Weyl structure is locally integrable so that the 
Weyl connection is locally equal to the Levi-Civita connection of some compatible metric. \\

If $(\gm,\psi)$ belongs to the Weyl structure and if $\nabla$ denotes the Riemann connection for $\gm$, the difference 
$\Wnabla-\nabla$ is a tensor $\Psi$ of type $(^1_2)$ such that for any vector fields $X$ and $Y$ and any 1-form $\theta$ 
(see \cite{HIGA,SCHOLZ1,SCHOLZ2,WEYL2}): 
\small
\begin{align}
\label{WnablaXY}
\Wnabla_XY &= \nabla_XY + \Psi_XY , & \Psi_XY &= \psi(X)Y + \psi(Y)X - \gm(X,Y) \psi_* , \\ 
\label{WnablaTheta}
\Wnabla_X \theta &= \nabla_X \theta - \Psi_X \theta , & \Psi_X \theta &= \psi(X)\theta + \theta (X) \psi - \theta(\psi_*) \gm(X) ,
\end{align}
\normalsize
where 
\footnote{$\gm$ is consi\-dered both as a bilinear form on the tangent bundle $TM$ and as a linear map from $TM$ to $T^*M$. 
The inverse mapping $\gm^{-1}$ form $T^*M$ to $TM$ is considered as a bilinear form on $T^*M$.} 
$\psi_*=\gm^{-1}(\psi)=g^{ij} \psi_i \partial_j$. \\
Let $(\gm, \psi) \equiv (\gm', \psi')$ according to (\ref{Weylequival}) with $\psi'=\psi+\phi$. 
If $\nabla'$ is the Levi-Civita connection associated to $\gm'$, (\ref{WnablaXY}) implies 
\small
\begin{align}
\label{nablanabla'}
\begin{split}
\nabla_XY - \nabla'_XY &= \phi(X)Y + \phi(Y)X - \gm(X,Y)\phi_*. 
\end{split}
\end{align}
\normalsize

The curvature tensor $\WRR$ of the Weyl structure is defined as in the Riemannian case by 
$\WRR_{XY}=[\Wnabla_X,\Wnabla_Y]-\Wnabla_{[X,Y]}$. 
$\WRR_{X,Y}$ is antisymmetric w.r.t. $X$ and $Y$ by construc\-tion. It satisfies the algebraic Bianchi identity and the 
differential Bianchi identity, which hold true for any torsion free connection (\cite{ROBBIN1}). 
The two other identities, Lie algebra invariance and permutation of pairs (see \cite{LEE} for instance), involve a metric 
and are satisfied by $\WRR$ if the Weyl structure is closed. 
The Ricci tensor has an intrinsic definition as a trace of the curvature tensor. However it may not be symmetric. 
T. Higa proved (\cite{HIGA}) that the Weyl-Ricci tensor $\WRRic$ of $\WRR$ is symmetric iff the Weyl 
structure is closed. 
\footnote{This follows from the identity $(\nabla_X\psi)(Y)-(\nabla_Y\psi)(X)=d\psi(X,Y)$
which applies for any torsion-free connection.} \\

For $(\gm,\psi)$ in the Weyl structure, we have the following relation between Ricci tensors 
(\cite{HIGA,MAEDER2017,ORNEA})
\small
\begin{align*}
\WRRic(X,Y)-\RRic(X,Y)&= (1-n)(\nabla_X\psi)(Y) + (\nabla_Y\psi)(X) - \gm(X,Y)\divpsi \\
&+ (n-2) \Big[ \psi(X)\psi(Y) - \gm^{-1}(\psi,\psi) \gm(X,Y) \Big] ,
\end{align*}
\normalsize
where $\RRic$ is the Ricci tensor for the Riemannian metric $\gm$ and $\divpsi$ 
is the divergence of $\psi_*=\gm^{-1}(\psi)$ computed with $\nabla$. 
As noted above, if $\psi$ is closed both Ricci tensors are symmetric and we get 
\small
\begin{align}
\label{WRicci}
\begin{split}
\WRRic-\RRic =& - \left[ (n-2) \gm^{-1}(\psi,\psi) +\divpsi \right] \gm +(n-2) 
\left(\psi\otimes\psi - \nabla\psi \right) , \\
\end{split} 
\end{align}
\normalsize
The scalar curvature is defined as the trace of the Ricci tensor. However $\RRic$ and $\WRRic$ are tensor of type $(^0_2)$ 
and the trace requires a metric. The trace of a tensor $T$ of type $(^0_2)$ for a given metric $\gm$ is 
$\Tr T=dx^i.\gm^{-1}(T(\partial_i))=g^{ij}T_{ij}$, where $T(\partial_i)$ is the 1-form $T(\partial_i,.)$.  
Then the scalar curvatures \small$\WRsc_\gm$ \normalsize of $\Wnabla$ and $\Rsc$ of $\nabla$, both computed with $\gm$, satisfy (see \cite{SCHOLZ2})
 \small
 \begin{align}
 \label{WScalar}
 \begin{split}
\WRsc_\gm - \Rsc = \Tr \left(\WRRic - \RRic \right) &= -(n-1)(n-2) \gm^{-1}(\psi,\psi) - 2(n-1) \divpsi \\
&= - (n-1) \left[ (n-2) \gm^{-1}(\psi,\psi) + 2 \, \divpsi \right]. 
\end{split}
\end{align}
\normalsize
The scalar curvature of $\Wnabla$ computed with a conformaly equivalent metric \small$\gm'=e^{2\Omega}\gm$ \normalsize is 
$\WRsc_{\gm'} = e^{-2\Omega}\,\WRsc_\gm$ so that a conformally invariant expression of the Weyl scalar curvature is 
the product $\WRsc_\gm \, \gm$. This justifies the following definition of the Weyl-Einstein tensor $\WGG$ as in the 
Riemannian case: 
\small
\begin{align}
\label{WGG}
\begin{split}
\WGG &= \WRRic -  \frac 1 2  \WRsc_\gm\,\gm . 
\end{split}
\end{align} 
\normalsize
The divergence of $\WGG$ vanishes when computed with the Weyl derivation, as a consequence of Poincar\'e lemma. 
Using (\ref{WRicci}) and (\ref{WScalar}) we get the following relation between the two tensors:
\small
\begin{align}
\label{WGGtensor}
\WGG &= \Gg + \tilde{\Lambda} \gm + (n-2) \left( \psi\otimes\psi- \nabla\psi \right) ,
\end{align} 
\normalsize
where $\tilde{\Lambda}$ is the scalar function defined by 
\small
\begin{align}
\label{Lambda}
\tilde{\Lambda}&=(n-2)\left[ \frac {(n-3)} 2 \gm^{-1}(\psi,\psi) +  \divpsi \right] .
\end{align} 
\normalsize
 The last term of (\ref{WGGtensor}) is a symmetric bilinear form since $\psi$ is closed. In the example of section 
\ref{example} the function $\tilde{\Lambda}$ diverges on the $z$ axis which carries the singularities of $\psi$, 
and tends to $0$ in perpendicular spatial directions. 

\section{Geodesics of Weyl connections }
\label{Weyl-geodesics}
A geodesic for the Weyl derivation $\Wnabla$ is a smooth curve $\lambda \mapsto \gamma(\lambda)$ on $M$  
such that  
\small
\begin{align}
\label{nabladgammadlambda}
\Wnabla_{\dlam{\gamma}} \dlam{\gamma} = \nabla_{\dlam{\gamma}} \dlam{\gamma} + 2\psi(\dlam{\gamma}) \dlam{\gamma} 
- \gm(\dlam{\gamma},\dlam{\gamma})\psi_* = 0 , 
\end{align} 
\normalsize
where $\nabla$ is the Levi-Civita connection of $\gm$. Due to (\ref{WnablaXY}) this holds for any $(\gm,\psi)$ in the 
Weyl structure. \\
Since $\Wnabla_\xi \gm = -2 \psi(\xi) \gm$, equation (\ref{nabladgammadlambda}) implies 
\begin{align}
\label{g-lambda}
\dlam{} \left[ \gm(\dlam{\gamma}, \dlam{\gamma}) \right] 
&=\Wnabla_{\dlam{\gamma}} \left[ \gm(\dlam{\gamma}, \dlam{\gamma}) \right] 
= -2\psi(\dlam{\gamma})\, \gm(\dlam{\gamma},\dlam{\gamma}) ,
\end{align}
so that $\gm(\dlam{\gamma},\dlam{\gamma}) $ is no more a constant. 

Poincar\'e lemma implies that for any point of $\gamma$ there exists a scalar function $\chi$ on $M$  such that 
$\psi = d\chi$ in a neighbourhood $U$ of that point. Define $\gm_{\chi} = e^{2\chi}\gm$ so that $(\gm_\chi,\psi-d\chi)$ 
belongs to the Weyl structure. 
The Levi-Civita connection $\nabla_\chi$ of $\gm_\chi$ and the Weyl connection $\Wnabla$ coincide in $U$.
 It follows that $\gamma$ is locally a geodesic of $\gm_\chi$ so that 
$\gm_\chi(\dlam{\gamma},\dlam{\gamma})$ is constant and we have 
\small
\begin{align}
\label{ggammagamma}
\begin{split}
\left. \gm(\dlam{\gamma},\dlam{\gamma})\right|_{\lambda} 
&= \left. \gm_\chi(\dlam{\gamma},\dlam{\gamma}))\right|_{\lambda}  e^{-2(\chi(\gamma_\lambda)-\chi(\gamma_{\lambda_0}))} ,
\end{split}
\end{align}
\normalsize
with the initial condition $\gm=\gm_\chi$ at $\gamma(\lambda_0)$. \\
With a suitable linear scaling of $\lambda$ we may assume that $\gm_\chi(\dlam{\gamma},\dlam{\gamma})=\kappa$ 
with $\kappa=-1$ for time-like geodesics, and setting $\lambda=c\tau$ defines the proper time $\tau$ on $\gamma$ 
(see \cite{DELHOM,ROMERO1,ROMERO4}). 
Furthermore, Fermi normal coordinates can be defined in $U$ for $\gm_\chi$, which insures Einstein equivalence principle. \\

The equation of a time-like geodesic $\gamma$ of $\Wnabla$ is a particular case of (\ref{nabladgammadlambda}): 
\small
\begin{align}
\label{time-like-0}
\Wnabla_{\dlam{\gamma}} \dlam{\gamma} = \nabla_{\dlam{\gamma}} \dlam{\gamma} + 2\psi(\dlam{\gamma}) \dlam{\gamma} 
= 0 , 
\end{align} 
\normalsize
A convenient change of parameter of $\gamma$ yields a solution $\tilde{\gamma}$ of the geodesic equations of $\gm$ 
so that $\gamma$ and $\tilde{\gamma}$ have the same shape. \\

A vector field $\xi$ is a Killing vector of $\gm$ if \small$L_\xi\gm=0$ \normalsize where $L$ is the Lie derivative. Then
\small$L_\xi\gm_{\chi} = 2 d\chi(\xi) \, \gm_{\chi}$ \normalsize so that $\xi$ is a {\it conformal} Killing vector of $\gm_\chi$.
For a torsion-free connection this is equivalent to \small$\gm(\nabla_X\xi,Y)+\gm(X,\nabla_Y\xi)=0$ \normalsize  
for any vector fields $X$ and $Y$. If $X=Y=\dlam{\gamma}$ this implies 
\small
\begin{align}
\label{WKilling-1}
\dlam{} \left( \gm_\chi(\xi,\dlam{\gamma}) \right) = \kappa \,d\chi(\xi). 
\end{align} 
\normalsize

The existence of closed and non-exact 1-forms on a manifold $M$ implies 
restrictions on its topology: The first de Rham cohomology group $H^1(M)$ must be non-trivial so that $M$ is neither simply 
connected nor contractible. A common model of spacetime is $M=\Real^{+}\times S^3$ 
where the spatial section is the unit sphere $S^3$ of $\Real^4$. A classical theorem states that the first cohomology 
group of $S^3$ is $H^1(S^3)=0$, and since $\Real^{+}$ is contractible, all closed 1-forms on $M$ are exact. %
The Poincar\'e dodecahedral space (see Luminet et al. \cite{LUM}) is a 3-dimensional compact manifold which is not 
simply-connected: its fundamental group is the binary icosahedral group $\textrm{I}^*$ of 120 elements, 
but it has the same homology as the sphere $S^3$ so that the first cohomology group is trivial. 

A simple method to obtain a non-trivial cohomology from $S^3$ is to remove a loop such as an embedding 
of $S^1$. It is shown in \cite{HATCHER} (proposition 2B.1) that if $h$ is an embedding of the sphere $S^k$ into $S^n$ 
with $k < n$, the singular homology group $H_i(S^n\smallsetminus h(S^k))$ is $\ZZ$ for $i=n-k-1$, and $0$ otherwise so that 
$H_1(S^3\smallsetminus h(S^1)) = \ZZ$. Then de Rham theorem states that the cohomology groups 
$H^1(S^3\smallsetminus h(S^1))$ is $\Real$: All closed 1-forms on $M=\Real^{+}\times (S^3\smallsetminus h(S^1))$ 
write $a\psi+d\ffi$ for some fixed closed 1-form $\psi$, with $a\in\Real$ and any smooth function $\ffi$. 

\section{A toy model}
\label{example}
We consider now the "manifold" \small$M=\Real^4\smallsetminus \Sigma$ \normalsize where 
$\Sigma =\{(ct,0,0,z)\,:\, t,z \in \Real\}$. 
The spatial sections of $M$ are $\Real^3 \smallsetminus \Delta$ where 
the $z$-axis $\Delta$ will carry the singularity of the 1-form $\psi$ and we clearly get 
$H^1(\Real^3\smallsetminus \Delta)=\Real$. \\
We use the Schwarzschild coordinates $(x^0,r,\theta,\ffi))$ where $x^0 = ct$ and  
$\gm$ denotes the Schwarzschild metric on $M$ with a Schwarzschild radius denoted $r_S$. 
The associated 1-form on $M$ is $\psi=\eps\, d\ffi$, 
where $\eps$ is a dimensionless constant, so that the divergence of $\psi_* = \frac{\eps}{(r\sin\theta)^2} \partial_\ffi$ is null. 
We consider the Weyl structure specified by the equivalence class of $(\gm,\psi)$. 

Equation (\ref{Lambda}) yields $\tilde{\Lambda} = \frac{\eps^2} {(r\sin\theta)^2}$ 
and the Weyl-Einstein tensor of equation (\ref{WGGtensor}) finally reads 
\small
\begin{align}
\begin{split}
\WGG &= \Gg + 
\frac{\eps^2} {(r\sin\theta)^2} \gm + 2\eps^2 \, d\ffi \otimes d\ffi \\
&+ 2\eps \left[ \frac 1 r \left( dr\otimes d\ffi+ d\ffi \otimes dr \right) 
+ \cot\theta \left( d\theta \otimes d\ffi + d\ffi \otimes d\theta \right) \right] .
\end{split}
\end{align}
\normalsize

The equation of a geodesic $\gamma$ of $\Wnabla$ follows from (\ref{nabladgammadlambda}): 
\small
\begin{align}
\label{Wnabla}
&\Wnabla_{\dlam{\gamma}} \dlam{\gamma} = \nabla_{\dlam{\gamma}}\dlam{\gamma} 
+ \eps \Bigg[ 2 \,\dlam{\ffi}\, \dlam{\gamma}
-\tfrac{ \gm\big(\dlam{\gamma},\dlam{\gamma}\big)} { (r\sin\theta)^2}  \partial_\ffi \Bigg] = 0 ,
\end{align} 
\normalsize
where $\nabla$ is the Levi-Civita connection of the Schwarzschild metric. \\

If $\gamma$ is a solution of the above equation, the coordinate $\gamma^\ffi$ sets a 
particular determination of $\ffi(\textrm{mod}\, 2\pi)$ on a simply-connected neighbourhood $U \subset M$ of $\gamma$. 
Then  $\gm_\chi = e^{2\eps \ffi } \gm$ is well defined on $U$ and the equivalence condition between $(\gm,\psi)$ and 
$(\gm_\chi,0)$ also holds on $U$, i.e. on the whole geodesic $\gamma$. 
This implies $\Wnabla = \nabla_\chi$, where $\nabla_\chi$ is the connection of $\gm_\chi$ and 
$\WGG$ is equal to the Einstein tensor $\Gg_\chi$ of $\gm_\chi$ on $U$. 
As noted in the previous section $\gm_\chi(\dlam{\gamma},\dlam{\gamma})$ is a constant that can be set to
 $-1$ for time-like geodesics. It follows that $\gm(\dlam{\gamma},\dlam{\gamma})= -e^{-2\eps \gamma^\ffi}$. 
 If $\gamma$ is a light-like geodesic the normalisation of $\dlam{\gamma}$ is reduced to  
$\gm(\dlam{\gamma},\dlam{\gamma})=0$. 

\subsection{Conformal Killing vectors}
As $\gm_\chi=e^{2\eps\ffi}\gm$ is independent of $x^0$, $\partial_{x^0}$ is a Killing vector and equation (\ref{WKilling-1}) 
implies 
\small
\begin{align*}
\begin{split}
\gm_\chi(\partial_{x^0},\dlam{\gamma}) &= e^{2\eps\ffi} \gm(\partial_{x^0},\dlam{\gamma}) 
= - e^{2\eps\ffi} \dlam{x^0} (1-\tfrac {r_S}r ) = - \engie_0 
\end{split}
\end{align*} 
\normalsize
where $\engie_0$ is a constant. \\
Since $\partial_\ffi$ is a Killing vector of $\gm$ it's a conformal Killing vector of $\gm_\chi$ with 
$L_{\partial_\ffi}\gm_\chi = 2\eps \gm_\chi$.  
Integration of equation (\ref{WKilling-1})  
is straightforward and we can set 
\small
\begin{align}
\label{LcalDef}
\begin{split}
\Lcal &= e^{2\eps\ffi} r^2\sin^2\theta \dlam{\ffi} = \Lcal_0 + \eps \kappa(\lambda-\lambda_0) , 
\end{split}
\end{align} 
\normalsize
where $\Lcal_0$ is fixed by initial conditions. 
The equations of geodesics write 
\small
\begin{align}
\label{Eqn10}
\begin{split}
\dlam{x^0} &= \engie_0  e^{-2\eps\ffi} (1-\tfrac {r_S}r )^{-1} , \\
\ddlam{r} &= \kappa \tfrac{r_S}{2r^2} e^{-2\eps\ffi} + r(1 - \tfrac{3r_S}{2r}) \left[(\dlam{\theta})^2 +
 \sin^2\theta (\dlam{\ffi})^2 \right] - 2\eps \dlam{r}\dlam{\ffi} , \\
 \ddlam{\theta} &= - \tfrac 2 r \dlam{r}\dlam{\theta} + \sin\theta\cos\theta (\dlam{\ffi})^2 - 2 \eps \dlam{\theta}\dlam{\ffi} , \\
\dlam{\ffi} &= e^{-2\eps\ffi} \tfrac{\Lcal_0 + \eps \kappa(\lambda-\lambda_0)}{r^2\sin^2\theta} .
\end{split}
\end{align} 
\normalsize 
In the case $r_S=0$, $\gm$ is invariant w.r.t. translations on the $z$-axis so that $\partial_z$ is a Killing vector of $\gm$. 
Since $d\chi (\partial_z) = 0$ it follows from (\ref{WKilling-1}) that $ \engie_z=\gm_\chi(\partial_z,\dlam{\gamma})$ is 
constant on $\gamma$. 
With $\partial_z=\cos\theta \,\partial_r - \tfrac{\sin\theta}r \, \partial_\theta$ it follows that 
\small
\begin{align}
\label{dlamgammaz}
 \dlam{\gamma^z} = \cos\theta \dlam{r} - r \sin\theta \dlam{\theta} = \gm(\partial_z,\dlam{\gamma}) 
 = e^{-2\eps\ffi}  \engie_z .
\end{align}
\normalsize

\subsection{Planar time-like geodesics}
\label{Planar}
\begin{figure}[H]
\drawfigtwo{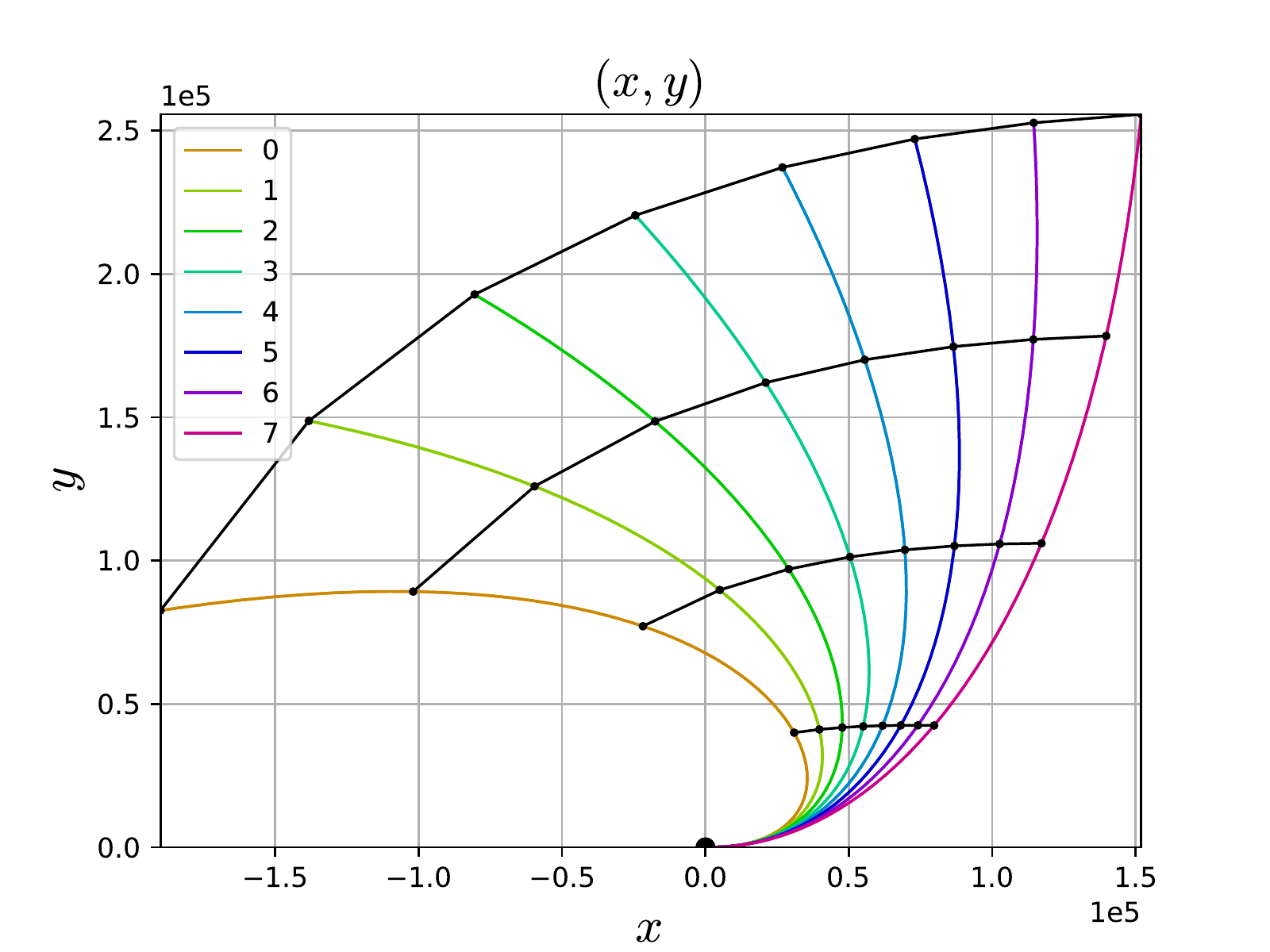}{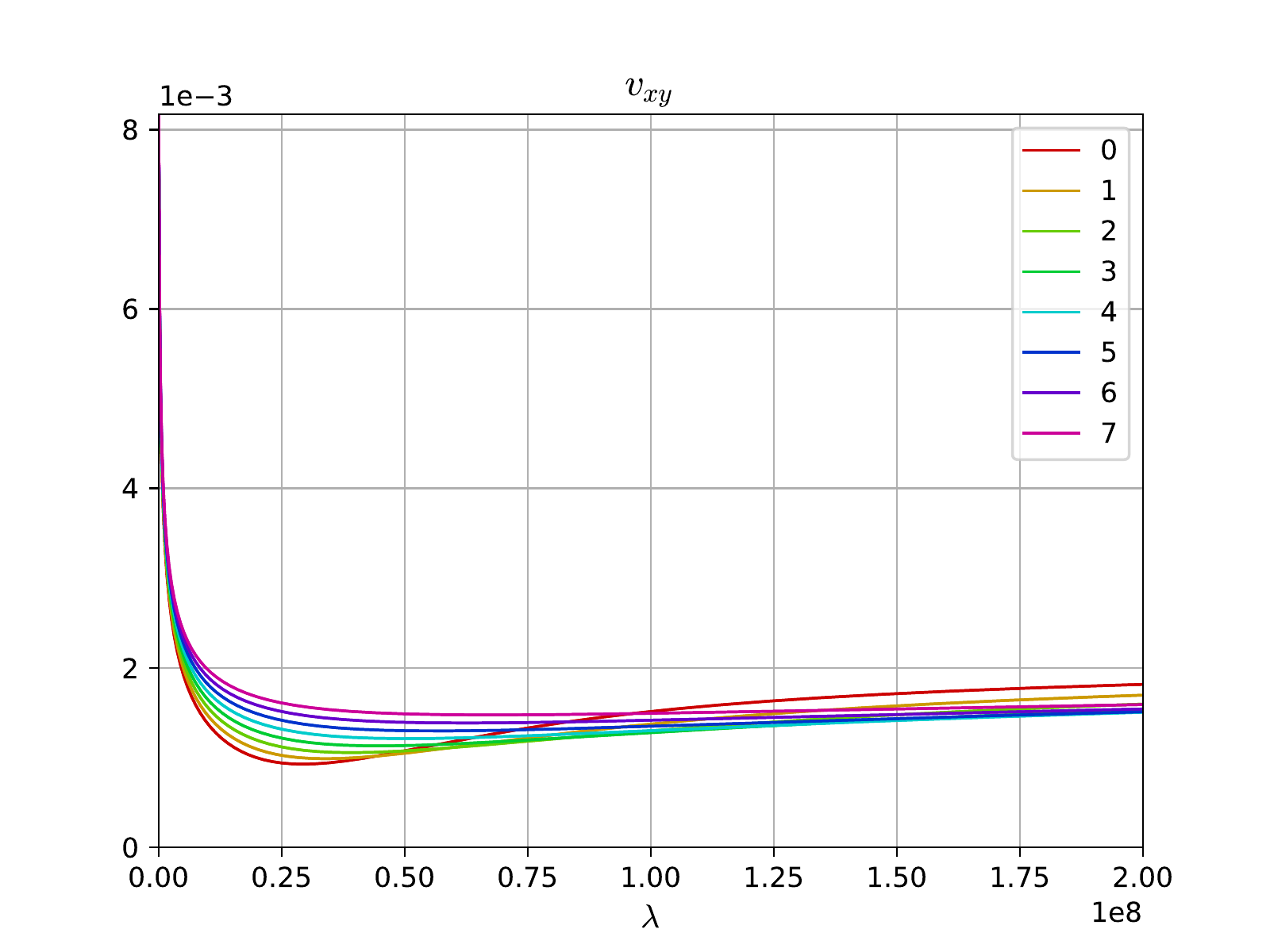}{0.39}
\caption{(Left) Black lines link planar geodesics at equal proper times. (Right) 2-d (total) velocities in the $xy$-plane 
as functions of proper times.}
\label{fig:rs>0}
\end{figure}

The eight simulations of geodesics presented in Fig.\ref{fig:rs>0} are solutions of (\ref{Eqn10}) in the symmetry plane $z=0$, 
with the Schwarzschild radius $r_S=0.1$ and $\eps=-10^{-6}$.  
Initial radial velocities are close to the escape velocity at $r_0=1.5\times 10^3$ and range linearly from 
\small$\dlam{r}|_0 = 0.008002$ \normalsize to \small$\dlam{r}|_0 = 0.008173$ \normalsize with index $0$ to $7$. 
The right side of Fig.\ref{fig:rs>0} shows that total velocities, including a radial component, are almost constant for proper 
times $\lambda> 0.5\times 10^8$.
 
\subsection{Case $r_S=0$}
\label{r_S=0}
This case reveals the effects of the 1-form $\eps\, d\ffi$ is the absence of curvature due to masses. 
The geodesic equations written in cylindrical coordinates \small$(x^0,\rho,\ffi,z)$ \normalsize show that for a geodesic 
$\gamma$ the $z$-coordinate satisfies \small$\dlam{\gamma^z} = \dlam{\gamma^z}\big|_0 e^{-2\eps\ffi}$\normalsize, 
a result which follows from the existence of the Killing vector $\partial_z$ as in (\ref{dlamgammaz}). 

\begin{figure}[H]
\drawfigtwo{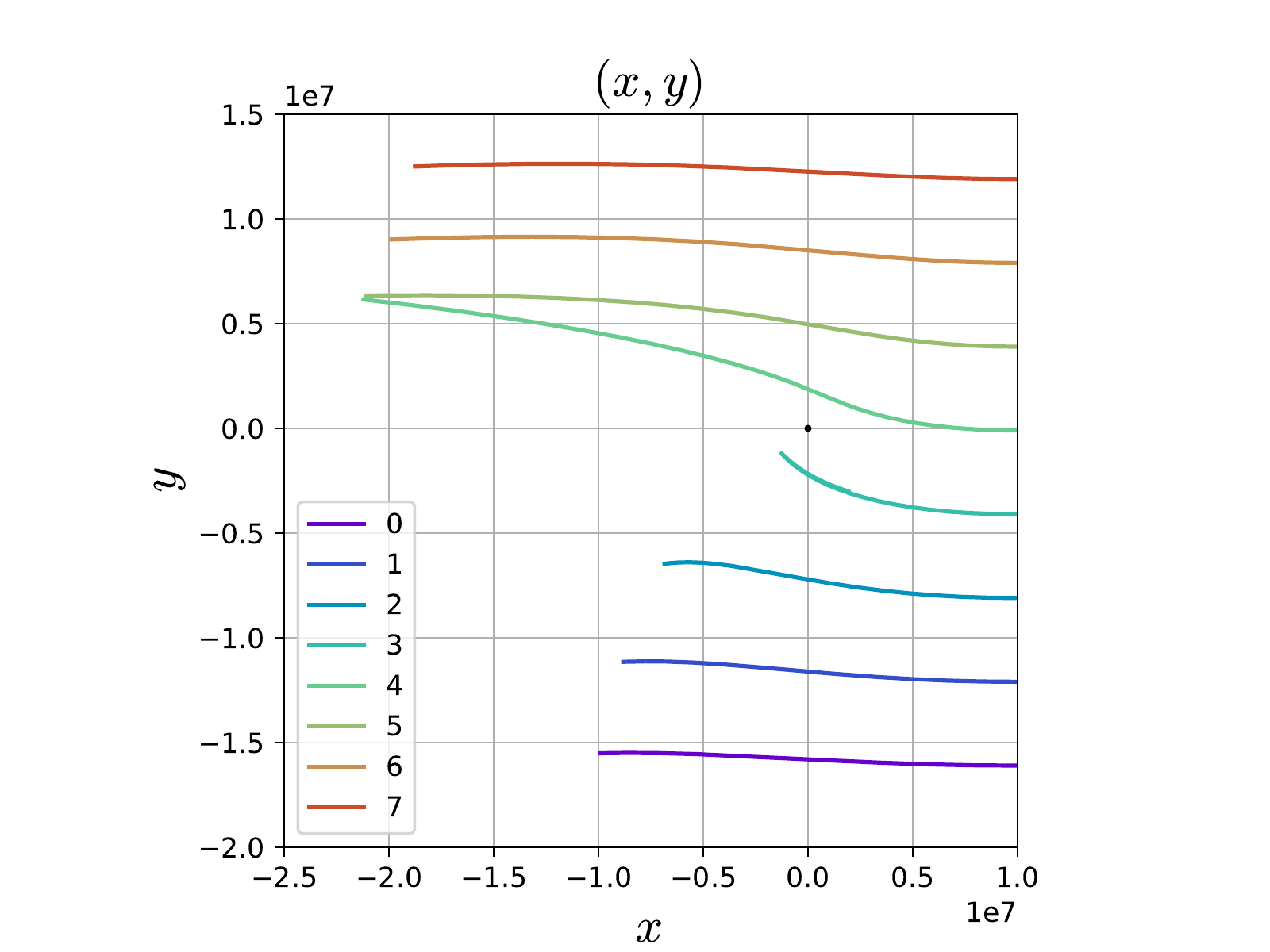}{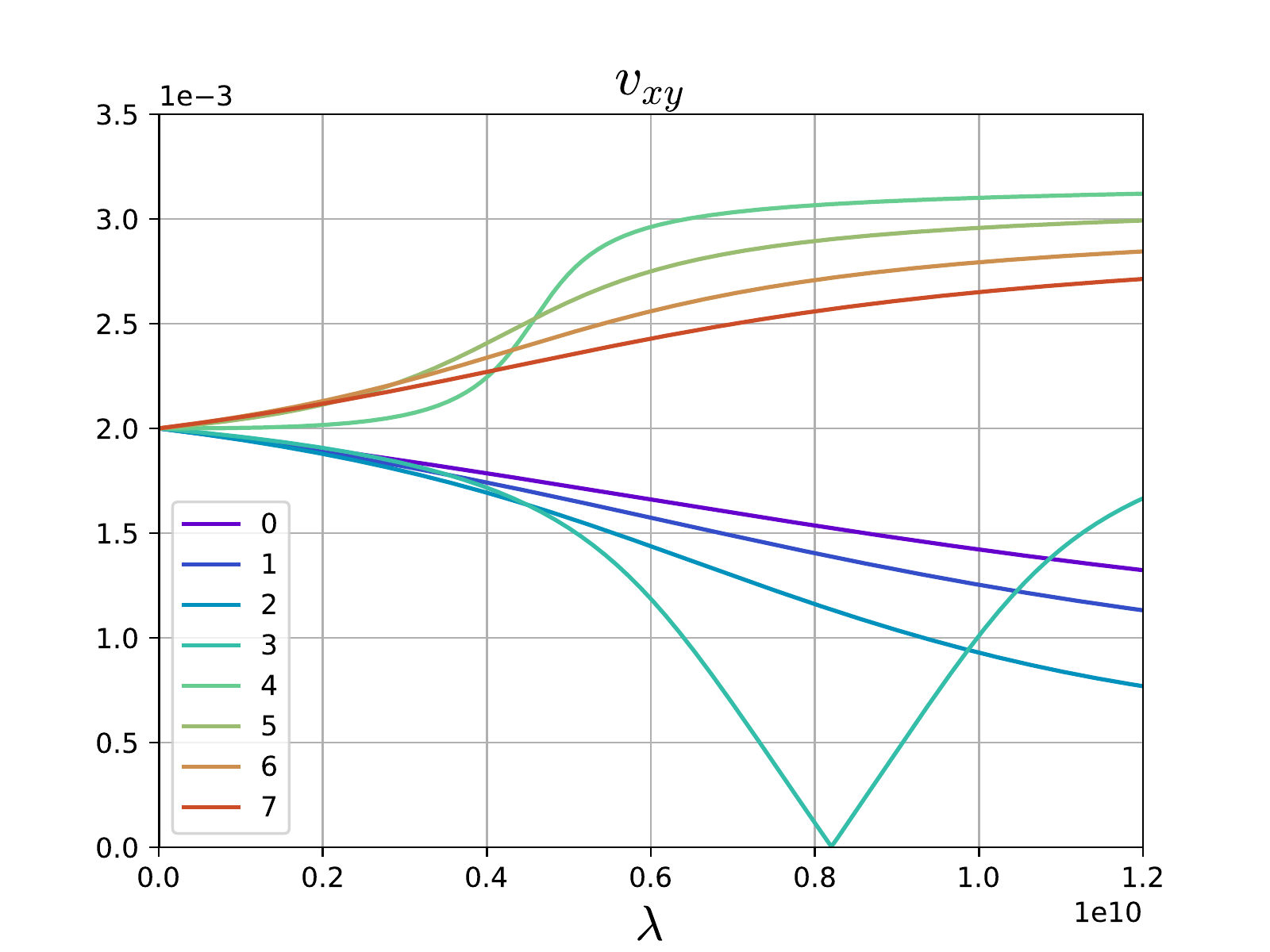}{0.4}
\caption{(Left) According to initial values geodesics are accelerated or slowed down. 
(Right) Total 2-d velocities as functions of proper time. The fourth geodesic is close to a cusp.}
\label{fig:rs=0}
\end{figure}

\begin{figure}[H]
\drawfigtwo{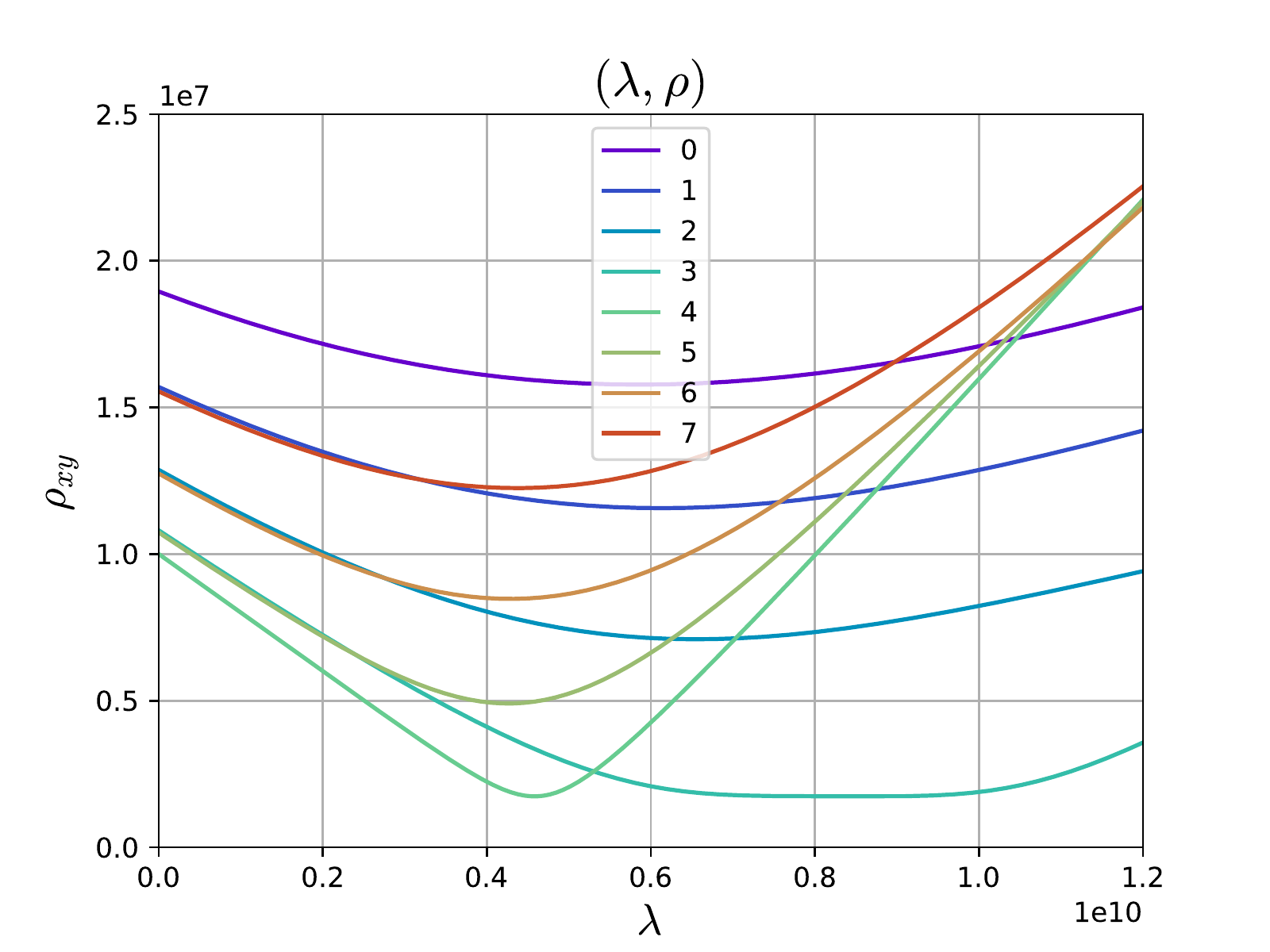}{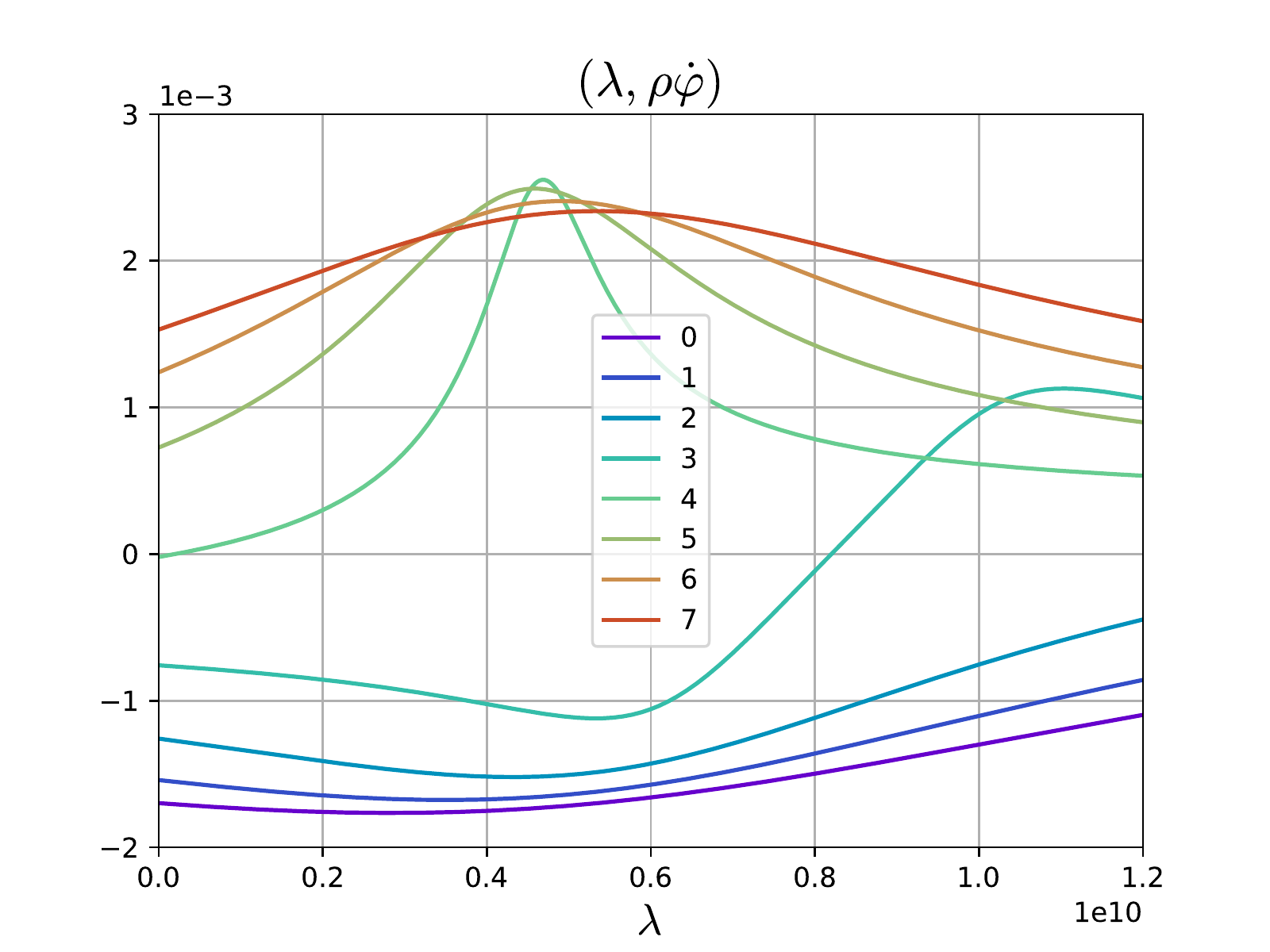}{0.4}
\caption{(Left) The geodesic of index 3 "stands" a longer time in the neighbourhood of the singularity. 
(Right) Tangential velocities as functions of proper time (\small$\dot{\ffi} = \dlam{\ffi}$\normalsize).}
\label{fig:rs=0_1}
\end{figure}

The eight simulations of geodesics of Fig.\ref{fig:rs=0} and Fig.\ref{fig:rs=0_1} were obtained for initial values 
$x_0 = 10^7$, $\dlam{x^0}= -0.002$ and $y_0$ values range linearly from $-1.61\times 10^7$ to $1.19\times 10^7$ 
for index $0$ to $7$. 
The two figures clearly show that geodesics are accelerated or slowed down according to which side of the singularity they go. 
We also see that the geodesic of index 3 "spends" more time at slow velocity in the neighbourhood of the $z$ axis.

\begin{figure}[H]
\drawfigtwo{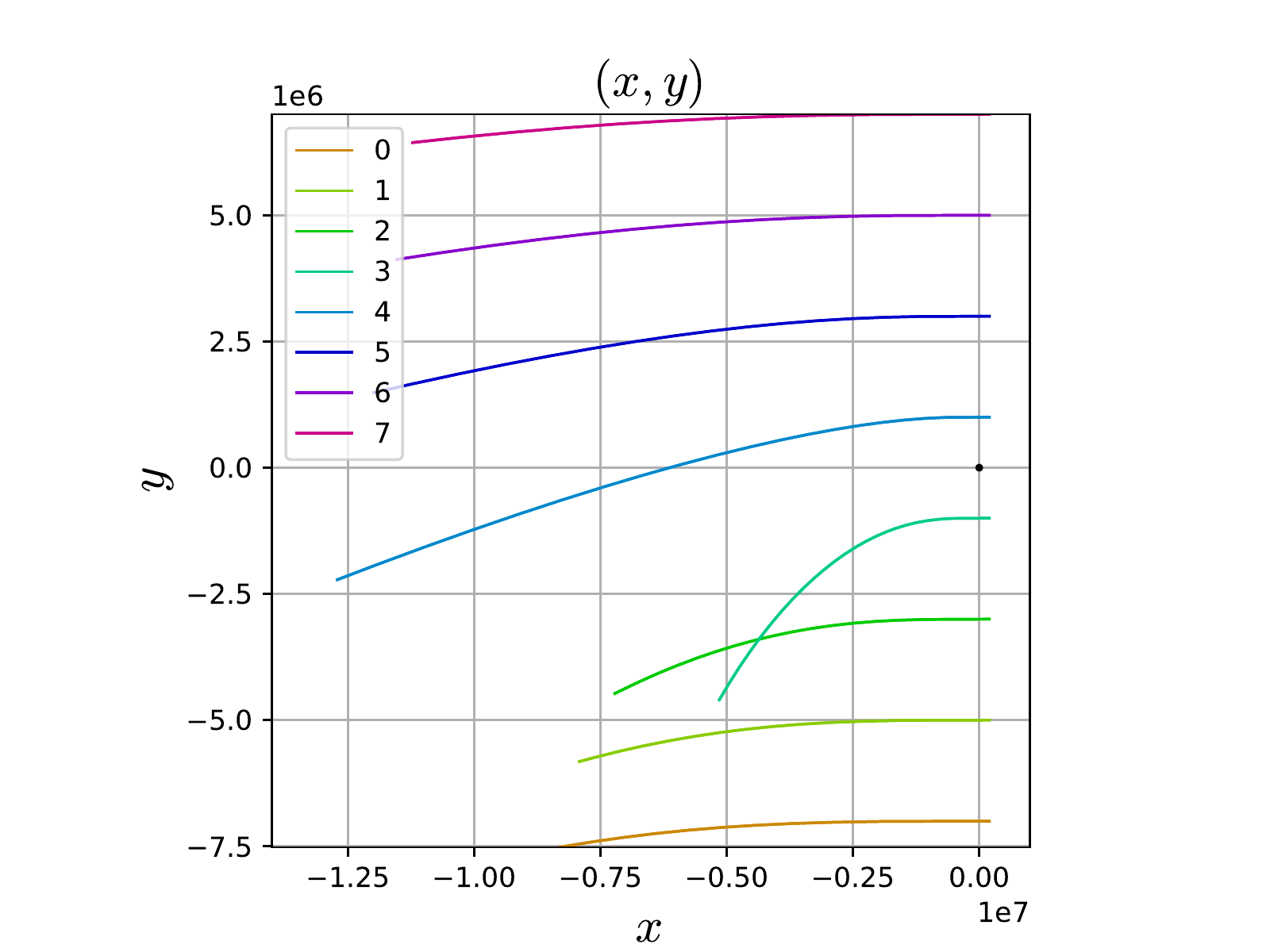}{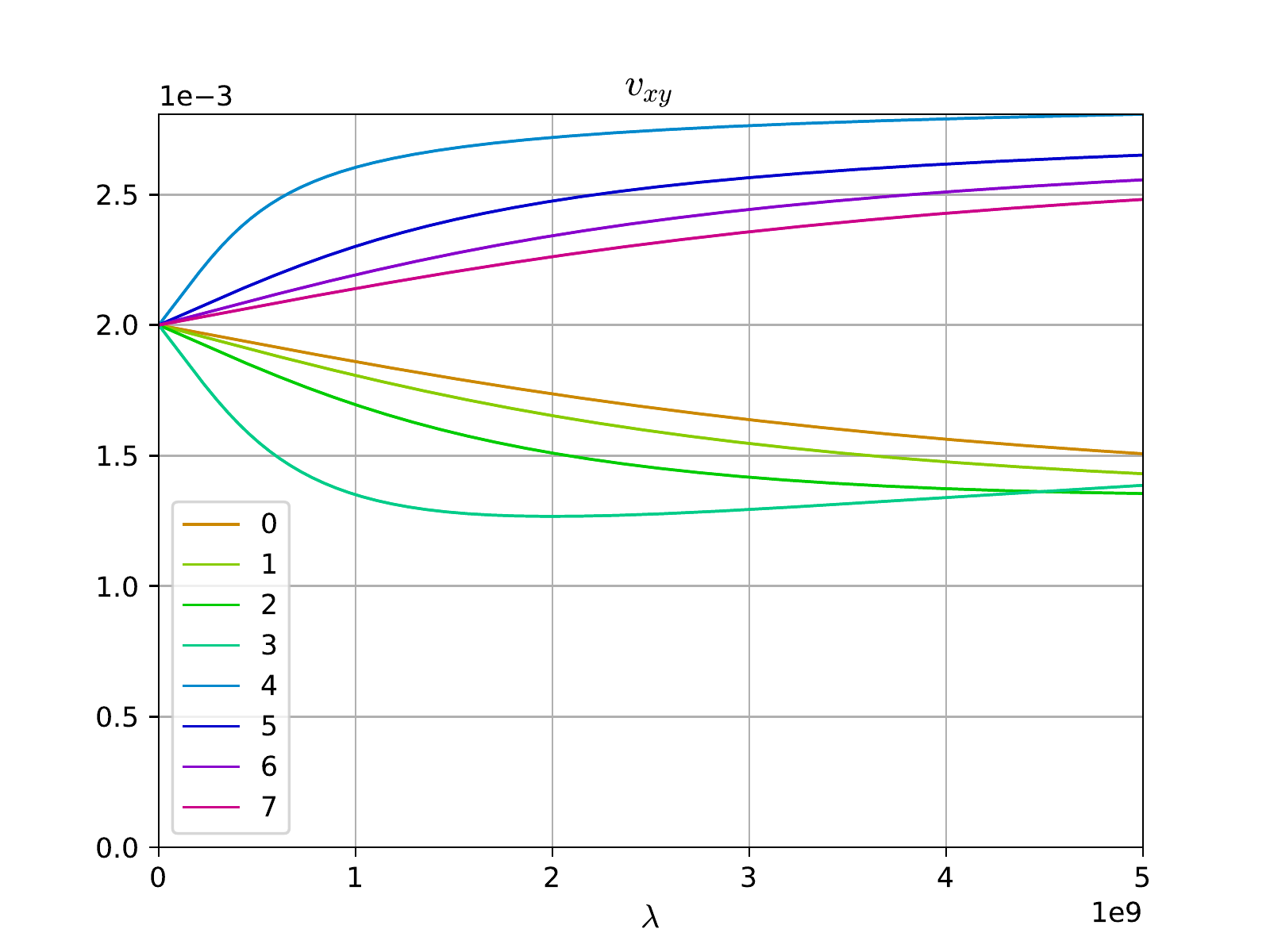}{0.41}
\caption{According to initial values geodesics are accelerated or slowed down. Right: 2-d velocities as functions of proper time.}
\label{fig:rs=0_2}
\end{figure}

Initial values for Fig.\ref{fig:rs=0_2} are $x_0 = 2\times 10^5$, $\dlam{x^0}= -0.002$ and $y_0$ values range linearly from 
$-7\times 10^6$ to $7\times 10^6$ for index $0$ to $7$. 
These figures also show that geodesics are accelerated or slowed down according to which side of the singularity they go. 
Notice that the shifts have the same sign if the initial conditions are rotated by $\pi$. 
These simulations for $r_S=0$ could be considered in connection with observations of cosmic filaments spin as reported in 
(\cite{TEMPEL,WANG}) and in (\cite{ALEXANDER}) where bimodal distributions of rotation speeds are observed. 
Likewise, the involvement of a linear singularity by a non-exact 1-form could suggest a connection with the Giant Arc 
recently discovered (\cite{LOPEZ}).

\subsection{Time-like geodesics}
A time-like geodesic $\gamma$ of $\Wnabla=\nabla_\chi$ is a solution of (\ref{time-like-0}). If a curve $\tilde{\gamma}$ 
is defined by \small$\gamma(\lambda) = \tilde{\gamma}(\mu(\lambda))$ \normalsize, where $\mu$ is a smooth strictly 
increasing function, then 
\small
\begin{align*}
\Wnabla_{\dlam{\gamma}} \dlam{\gamma} 
&= (\dlam{\mu})^2 \nabla_{\dmu{\tilde{\gamma}}} \dmu{\tilde{\gamma}} +  \ddlam{\mu} \dmu{\tilde{\gamma}} 
+ 2 \eps \dlam{\gamma^\ffi} \dlam{\mu}\, \dlam{\tilde{\gamma}} = 0 ,
\end{align*} 
\normalsize
and $\tilde{\gamma}$ is a geodesic of $\gm$ iff $\mu$ satisfies the equation 
\small$\ddlam{\mu} + 2 \eps \dlam{\gamma^\ffi} \dlam{\mu} =0$ \normalsize . \\

Let $a_0$ and $a_1$ be two points 
of $\gamma$. The derivation of Einstein effect states that a periodic signal of frequency $\nu_0$ emitted at 
$a_0$ is received at $a_1$ with frequency $\nu_1$ such that 
\small
\begin{align*}
\frac{\nu_1}{\nu_0} &= \left( \frac{\gm_\chi^{\,00}(a_0)}{\gm_\chi^{\,00}(a_1)} \right)^{1/2} 
= e^{-\eps(\ffi_1-\ffi_0)} \left( \frac{1-\tfrac{r_S}{r_0}}{1-\tfrac{r_S}{r_1}} \right)^{1/2} . 
\end{align*} 
\normalsize
In the limit \small$r_S=0$ \normalsize the geodesics are straight lines and if they are not parallel to the $z$ axis 
we can choose $a_0$ and $a_1$ large enough such that $|\ffi_1-\ffi_0| \simeq \pi$. If $\ffi_1>\ffi_0$ the geodesic is counterclockwise 
and \small$\nu_1/\nu_0 \simeq e^{-\eps\pi}$\normalsize, or clockwise in the other case and 
\small$\nu_1/\nu_0 \simeq e^{\eps\pi}$. \normalsize
As in section \ref{r_S=0} the frequency shifts depend on which side of the singularity the geodesics go and they have the same sign as the Doppler effects associated to the time-like geodesics. 

Assume now that two light sources of equal wave length are located at \small$r_0>r_S$\normalsize, \small$\theta=0$ \normalsize 
and \small$\ffi_\pm = \pm \pi/2$\normalsize. 
An observer at \small$r_1\gg r_0$ \normalsize, \small$\theta=0$ \normalsize and \small$\ffi=\pi$ \normalsize will receive two 
different frequencies \small$\nu_\pm$ \normalsize such that \small$\nu_{+}/\nu_{-} \approx e^{-\eps\pi}$. \normalsize

\section{Conclusions}
The integrable Weyl conformal geometry raised a great interest during the 20$^{th}$ century and in the past years. One of the reasons for this interest relies on the open issues in astrophysics and in cosmology, which gave rise to alternative theories 
of gravitation. In this paper we have considered the case of locally integrable (non-exact) Weyl conformal geometry, 
perhaps the simplest extension of Einstein's theory of general relativity. 
The first point relates to the topological constraints which exclude simply connected spacetime manifolds and imply 
linear singularities. 
The main result is the Weyl-Einstein tensor which includes a function in place of the cosmological constant $\Lambda$. 

The toy model of last section is based on a Schwarzschild metric and therefore it is far from addressing main issues of 
astrophysics. However this simple example suggests that the Weyl locally integrable conformal gravity provides a 
slightly different starting point for both rotations curves of spiral galaxies 
and suggests a topological basis for cosmic filament. 

I would like to thank E. Scholz for fruitful discussions, for his support and his encouragements.

\bibliography{WLICG.bbl}

\begin{thebibliography}{10}

\bibitem{ALEXANDER}
S.~Alexander, C.~Capanelli, E.~G.~M. Ferreira, and E.~McDonough.
\newblock Cosmic filament spin from dark matter vortices.
\newblock 2021.
\newblock URL: \url{https://arxiv.org/abs/2111.03061}, \href
  {https://doi.org/10.48550/ARXIV.2111.03061}
  {\path{doi:10.48550/ARXIV.2111.03061}}.

\bibitem{CALDERBANK}
D.~M.~J. {Calderbank} and H.~G. {Pedersen}.
\newblock {Einstein-Weyl geometry.}
\newblock In {\em {Surveys in differential geometry. Vol. VI: Essays on
  Einstein manifolds. Lectures on geometry and topology.}}, pages 387--423.
  Cambridge, MA: International Press, 1999.

\bibitem{DELHOM}
Adri{\`{a}} Delhom, Iarley~P. Lobo, Gonzalo~J. Olmo, and Carlos Romero.
\newblock Conformally invariant proper time with general non-metricity.
\newblock {\em The European Physical Journal C}, 80(5), may 2020.
\newblock URL: \url{https://doi.org/10.1140%2Fepjc%2Fs10052-020-7974-y}, \href
  {https://doi.org/10.1140/epjc/s10052-020-7974-y}
  {\path{doi:10.1140/epjc/s10052-020-7974-y}}.

\bibitem{DELIDUMAN}
Cemsinan Deliduman, O{\u{g}}uzhan Ka{\c{s}}{\i}k{\c{c}}{\i}, and Bar{\i}{\c{s}}
  Yap{\i}{\c{s}}kan.
\newblock Flat galactic rotation curves from geometry in weyl gravity.
\newblock {\em Astrophysics and Space Science}, 365(3), mar 2020.
\newblock URL: \url{https://doi.org/10.1007%2Fs10509-020-03764-y}, \href
  {https://doi.org/10.1007/s10509-020-03764-y}
  {\path{doi:10.1007/s10509-020-03764-y}}.

\bibitem{EDERY2}
A.~Edery, A.~A. M{\'{e}}thot, and M.~B. Paranjape.
\newblock {LETTER}: Gauge choice and geodetic deflection in conformal gravity.
\newblock {\em General Relativity and Gravitation}, 33(11):2075--2079, nov
  2001.
\newblock URL: \url{https://doi.org/10.1023%2Fa%3A1013011312648}, \href
  {https://doi.org/10.1023/a:1013011312648}
  {\path{doi:10.1023/a:1013011312648}}.

\bibitem{EDERY1}
A.~Edery and M.~B. Paranjape.
\newblock Classical tests for weyl gravity: Deflection of light and time delay.
\newblock {\em Physical Review D}, 58(2), jun 1998.
\newblock URL: \url{https://doi.org/10.1103%2Fphysrevd.58.024011}, \href
  {https://doi.org/10.1103/physrevd.58.024011}
  {\path{doi:10.1103/physrevd.58.024011}}.

\bibitem{FOLLAND}
G.~B. Folland.
\newblock Weyl manifolds.
\newblock {\em J. Differential Geom.}, 4(2):145--153, 1970.
\newblock \href {https://doi.org/10.4310/jdg/1214429379}
  {\path{doi:10.4310/jdg/1214429379}}.

\bibitem{HATCHER}
A.~Hatcher.
\newblock {\em Algebraic Topology}.
\newblock Cambridge University Press, 2002.
\newblock URL: \url{http://pi.math.cornell.edu/~hatcher/}.

\bibitem{HIGA}
T.~Higa.
\newblock {Weyl Manifolds and Einstein-Weyl Manifolds}.
\newblock In {\em Commentarii Mathematici Sancti Pauli}, volume~42, 1993.

\bibitem{LEE}
J.~M. Lee.
\newblock {\em {Riemannian manifolds. An introduction to curvature}}, volume
  176 of {\em Graduate Texts in Mathematics}.
\newblock Springer, 1997.

\bibitem{LOPEZ}
Alexia~M. Lopez, Roger~G. Clowes, and Gerard~M. Williger.
\newblock A giant arc on the sky, 2022.
\newblock URL: \url{https://arxiv.org/abs/2201.06875}, \href
  {https://doi.org/10.48550/ARXIV.2201.06875}
  {\path{doi:10.48550/ARXIV.2201.06875}}.

\bibitem{LUM}
J.-P. Luminet.
\newblock {The Status of Cosmic Topology after Planck Data}.
\newblock {\em {Universe}}, 2(1):1, 2016.
\newblock URL: \url{https://hal.archives-ouvertes.fr/hal-01291848}, \href
  {https://doi.org/10.3390/universe2010001}
  {\path{doi:10.3390/universe2010001}}.

\bibitem{MAEDER2017}
A.~Maeder.
\newblock An alternative to the {$\Lambda$}cdm model: The case of scale
  invariance.
\newblock {\em The Astrophysical Journal}, 834(2):194, Jan 2017.
\newblock \href {https://doi.org/10.3847/1538-4357/834/2/194}
  {\path{doi:10.3847/1538-4357/834/2/194}}.

\bibitem{MAEDER2019_2}
A.~Maeder and V.~G. Gueorguiev.
\newblock Scale-invariant dynamics of galaxies, mond, dark matter, and the
  dwarf spheroidals.
\newblock {\em Monthly Notices of the Royal Astronomical Society},
  492(2):2698--2708, Dec 2019.
\newblock URL: \url{http://dx.doi.org/10.1093/mnras/stz3613}, \href
  {https://doi.org/10.1093/mnras/stz3613} {\path{doi:10.1093/mnras/stz3613}}.

\bibitem{MANNHEIM_1995}
P.~D. Mannheim.
\newblock Cosmology and galactic rotation curves, 1995.
\newblock \href {http://arxiv.org/abs/astro-ph/9511045}
  {\path{arXiv:astro-ph/9511045}}.

\bibitem{MANNHEIM_2006}
P.~D. Mannheim.
\newblock Alternatives to dark matter and dark energy.
\newblock {\em Progress in Particle and Nuclear Physics}, 56(2):340--445, Apr
  2006.
\newblock \href {https://doi.org/10.1016/j.ppnp.2005.08.001}
  {\path{doi:10.1016/j.ppnp.2005.08.001}}.

\bibitem{MANNHEIM_2012}
P.~D. Mannheim and J.~G. O'Brien.
\newblock Fitting galactic rotation curves with conformal gravity and a global
  quadratic potential.
\newblock {\em Phys. Rev. D}, 85:124020, Jun 2012.
\newblock \href {https://doi.org/10.1103/PhysRevD.85.124020}
  {\path{doi:10.1103/PhysRevD.85.124020}}.

\bibitem{ORNEA}
L.~Ornea.
\newblock {\em {Weyl structures in quaternionic geometry. A state of the art}},
  volume~1.
\newblock Univ. degli Studi della Basilicat, 2002.
\newblock \href {http://arxiv.org/abs/math/0105041}
  {\path{arXiv:math/0105041}}.

\bibitem{ROBBIN1}
J.~W. Robbin and D.~A. Salamon.
\newblock {\em Introduction to differential geometry}, 2018.
\newblock URL:
  \url{https://people.math.ethz.ch/~salamon/PREPRINTS/diffgeo.pdf}.

\bibitem{ROMERO1}
Carlos Romero.
\newblock Is weyl unified theory wrong or incomplete?, 2015.
\newblock URL: \url{https://arxiv.org/abs/1508.03766}, \href
  {https://doi.org/10.48550/ARXIV.1508.03766}
  {\path{doi:10.48550/ARXIV.1508.03766}}.

\bibitem{ROMERO4}
T.~A.~T. Sanomiya, I.~P. Lobo, J.~B. Formiga, F.~Dahia, and C.~Romero.
\newblock Invariant approach to weyl's unified field theory.
\newblock {\em Phys. Rev. D}, 102:124031, Dec 2020.
\newblock URL: \url{https://link.aps.org/doi/10.1103/PhysRevD.102.124031},
  \href {https://doi.org/10.1103/PhysRevD.102.124031}
  {\path{doi:10.1103/PhysRevD.102.124031}}.

\bibitem{SCHOLZ3}
E.~Scholz.
\newblock {Weyl geometry in late 20th century physics}, 11 2011.
\newblock URL:
  \url{http://www.weylmann.com/Weyl\%20Geometry\%20in\%20Late\%2020th\%20Century.pdf},
  \href {http://arxiv.org/abs/1111.3220} {\path{arXiv:1111.3220}}.

\bibitem{SCHOLZ1}
E.~Scholz.
\newblock {\em Paving the Way for Transitions---A Case for Weyl Geometry},
  pages 171--223.
\newblock Springer New York, New York, NY, 2017.
\newblock URL:
  \url{http://philsci-archive.pitt.edu/10889/4/scholz_paving_2014_07.pdf},
  \href {https://doi.org/10.1007/978-1-4939-3210-8}
  {\path{doi:10.1007/978-1-4939-3210-8}}.

\bibitem{SCHOLZ2}
E.~Scholz.
\newblock {\em The Unexpected Resurgence of Weyl Geometry in late 20th-Century
  Physics}, pages 261--360.
\newblock Springer New York, New York, NY, 2018.

\bibitem{TEMPEL}
{Tempel, E.}, {Kipper, R.}, {Saar, E.}, {Bussov, M.}, {Hektor, A.}, and {Pelt,
  J.}
\newblock Galaxy filaments as pearl necklaces.
\newblock {\em A\&A}, 572:A8, 2014.
\newblock \href {https://doi.org/10.1051/0004-6361/201424418}
  {\path{doi:10.1051/0004-6361/201424418}}.

\bibitem{WANG}
P.~Wang, N.~I. Libeskind, E.~Tempel, X.~Kang, and Q.~Guo.
\newblock Possible observational evidence that cosmic filaments spin.
\newblock {\em Nature Astronomy}, 2021.
\newblock URL: \url{https://www.nature.com/articles/s41550-021-01380-6}, \href
  {https://doi.org/10.1038/s41550-021-01380-6}
  {\path{doi:10.1038/s41550-021-01380-6}}.

\bibitem{WEYL1}
H.~Weyl.
\newblock {Gravitation und Elektrizit\"at}.
\newblock {\em Sitz. {K\"on}. Preuss. Akad. Wiss.}, pages 465--480, 1918.
\newblock URL:
  \url{http://neo-classical-physics.info/uploads/3/4/3/6/34363841/weyl_-_grav._and_electr.pdf}.

\bibitem{WEYL2}
H.~Weyl.
\newblock Reine {Infinitesimalgeometrie}.
\newblock {\em Mathematische Zeitschrift 2}, pages 384--411, 1918.
\newblock URL:
  \url{http://neo-classical-physics.info/uploads/3/4/3/6/34363841/weyl_-_pure_inf._geom..pdf}.

\end{thebibliography}
\bibliographystyle{plainurl}
\end{document}